\documentclass[journal]{IEEEtran}
\usepackage{blindtext, graphicx,booktabs}
       \usepackage{cite}
\usepackage{verbatim}
\usepackage{amsmath}
\usepackage{subfigure}

       \usepackage{amssymb}
       \allowdisplaybreaks[4]

\IEEEoverridecommandlockouts   
\ifCLASSINFOpdf

\else

\fi

\begin{document}
%
\title{Improving Localization Accuracy in Connected Vehicle Networks Using Rao-Blackwellized Particle Filters: Theory, Simulations, and Experiments}

\author{Macheng~Shen,
        Ding~Zhao,
        Jing~Sun 
        and Huei~Peng~
\thanks{*This work is funded by the Mobility Transformation Center at the University of Michigan with grant No. N021548. The first two authors, M. Shen and D. Zhao have equally contributed to this research.}%
\thanks{M. Shen and J. Sun are with the Department
of Naval Architecture and Marine Engineering, University of Michigan, Ann Arbor,
MI, 48109. (e-mail: macshen@umich.edu; jingsun@umich.edu)}
\thanks{D. Zhao (corresponding author, e-mail: zhaoding@umich.edu) is with the University of Michigan Transportation Research Institute, Ann Arbor, MI, 48109. }
 
 \thanks{H. Peng is with the Department of Mechanical Engineering, University of Michigan, Ann Arbor, MI, 48109. (e-mail: hpeng@umich.edu)}

\thanks{Manuscript received XXX XX, 2016; revised XXX XX, 201X.}}

\markboth{IEEE TRANSACTIONS ON INTELLIGENT TRANSPORTATION SYSTEMS,~Vol.~XX, No.~XX, December~2016}%
{Shell \MakeLowercase{\textit{et al.}}: Bare Demo of IEEEtran.cls for Journals}

\maketitle

\begin{abstract}
A crucial function for automated vehicle technologies is accurate localization. Lane-level accuracy is not readily available from low-cost Global Navigation Satellite System (GNSS) receivers because of factors such as multipath error and atmospheric bias. Approaches such as Differential GNSS can improve localization accuracy, but usually require investment in expensive base stations. Connected vehicle technologies provide an alternative approach to improving the localization accuracy. It will be shown in this paper that localization accuracy can be enhanced using crude GNSS measurements from a group of connected vehicles, by matching their locations to a digital map. A Rao-Blackwellized particle filter (RBPF) is used to jointly estimate the common biases of the pseudo-ranges and the vehicle positions. Multipath biases, which introduce receiver-specific (non-common) error, are mitigated by a multi-hypothesis detection-rejection approach. The temporal correlation of the estimations is exploited through the prediction-update process. The proposed approach is compared to existing methods using both simulations and experimental results. It was found that the proposed algorithm can eliminate the common biases and reduce the localization error to below 1 meter under open sky conditions.
\end{abstract}


%
\IEEEpeerreviewmaketitle

\section{Introduction}
\begin{figure}[htbp]
  \centering
  \includegraphics[width=2.5in]{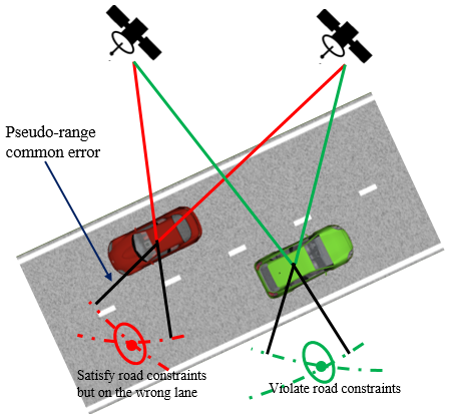}   
  \caption{Illustration of correlated GNSS localization error due to correlated pseudo-range error}
  \label{localization_bias}
\end{figure}

An essential function of intelligent transportation systems is accurate localization. A Global Navigation Satellite System (GNSS) receiver calculates its position from pseudo-range measurements of multiple satellites. Pseudo-ranges contain error which can be decomposed into common error (due to satellite clock error, ionospheric and tropospheric delays) and non-common error (due to receiver noise, receiver clock error and multipath error). The nominal accuracy of pseudo-ranges for a single-band receiver is about 10 to 20 meters, which results in a position error of several meters \cite{kaplan2005understanding}. Without further improvement, this crude GNSS error is too large for many safety functions as the lane of vehicles cannot be robustly identified. Fig.~\ref{localization_bias} shows the biased positioning caused by pseudo-range error.

\indent Differential GNSS (DGNSS) is an enhancement to GNSS that can achieve sub-meter level accuracy by correcting the common biases through a network of fixed reference stations. Moreover, centimeter-level accuracy is achievable by the Real Time Kinematic (RTK) technique, which uses carrier phase measurements to provide real-time corrections \cite{masumoto1993global}. These techniques, however, rely on an expensive infrastructure. In this paper, we explore an alternative low-cost solution for lane-level accurate localization by using only crude GNSS measurements from a set of connected vehicles.\\
\indent The following sections are arranged as follows. In Section 2, related works are reviewed. The challenges of localization using only single frequency market mass receivers are discussed. In Section 3, the derivation of the localization enhancement algorithm is presented in detail. Simulation results are presented in Section 4. The simulation scenario and error models are introduced, followed by performance analysis of the proposed method compared with existing methods. In Section 5, experimental results are presented using u-blox EVK-6T, an automotive grade receivers with automotive patch antenna. The implementation details are first described. Then the raw data is processed to verify the error models. Finally, the performance of the algorithms using the raw data is shown. Conclusions are presented in Section 6.
\section{Related Works}
The GNSS localization accuracy can also be improved using either sensor fusion or map matching, with some promising results shown in recent years \cite{shunsuke2015gnss,heidenreich2015laneslam,dominguez2015optimization}. \\
\indent Inertial Navigation System (INS) in combination with vehicle dynamics can be used to improve vehicle location estimation. Data fusion algorithms such as Extended Kalman Filter (EKF) \cite{toledo2007high}, Unscented Kalman Filter (UKF) \cite{zhang2005navigation} and particle filter \cite{caron2007particle} can be used to fuse GNSS and INS measurements, leading to high accuracy navigation solutions.\\
\indent With the rapid development of digital maps, navigation algorithms based on map matching have also been extensively studied \cite{toledo2007high,zhang2005navigation,white2000some,kim2016lane}. Map matching algorithms match the noisy GNSS positioning results to a trajectory that satisfies known road geometry constraints. Additional sensors such as cameras or lidars in combination with high-definition maps can reduce localization error down to the centimeter-level \cite{kim2016lane}. This approach requires both accurate sensors and accurate maps.  \\
\indent With the deployment of Dedicated Short Range Communications (DSRC) technique in the real world \cite{huang2016empirical}, connected vehicles provide an alternative for improving localization error by correcting common localization error of multiple GNSS receivers installed on multiple vehicles \cite{bezzina2014safety}. Alam et al. \cite{alam2013relative} developed a cooperative positioning method that improves the relative positioning between two vehicles by fusing the shared pseudo-range observations. The relative positioning accuracy of this method outperforms that of DGNSS alone. Wang \cite{wang2015cooperative} proposed an augment DGPS that tightly integrated DGPS, range and bearing observations in a small network of vehicles or infrastructure points, which outperforms the DGPS in terms of absolute positiong. An alternative to improving GNSS absolute positioning without incurring infrastructure costs is cooperative map matching (CMM). Assuming that most vehicles travel within lanes, the correction to the common localization biases can be determined so that the corrected positions of a group of vehicles best fit the map. Recently, the capability of CMM to mitigate the biases from the localization results obtained from low-cost GNSS receivers alone has been demonstrated in Rohani et al. \cite{rohani2016novel}. \\
\indent The three main difficulties arise for CMM using low-cost GNSS receivers alone are:

\begin{itemize}
 \item Effect of non-common error: Because of non-common error, a correction for the common localization error that makes all vehicle positions compatible with the road constraints may not exist. If the road constraints are enforced aggressively without considering the non-common error, the localization solution may overly converge, i.e., the variance of the estimation error may be underestimated. 
 \item Correlation between common biases and vehicle position estimation: the estimated common bias may induce error in the estimated vehicle positions and vice versa. If they are estimated sequentially, the data fusion scheme should be designed to avoid over-convergence due to fusion of correlated data. 
 \item Incorporation of road constraints: Lane position constraints of real roads cannot be easily described analytically. Incorporation of these constraints in map matching requires a flexible filter scheme. 
 \end{itemize}
\indent
\indent Wang et al. \cite{wang2013decentralized} presented a decentralizaed approach that uses local EKFs and an optimal global fusion scheme to deal with the correlation. Nonetheless, their global fusion approach requires a full order covariance matrix inversion and no road constraints are considered in their work. Rohani et al. \cite{rohani2016novel} presented a particle-based CMM algorithm to address the three difficulties mentioned above. The first difficulty was addressed by a weighted road map approach to preserve consistency. The second difficulty is handled by tracking the origin of the common bias corrections from different vehicles and fusing only those corrections from independent sources to avoid data incest, which avoids over-convergence. In their approach, some of the correlated corrections containing additional information have to be discarded. The third difficulty is handled by a particle-based approach that uses only the vehicle position estimation of the current epoch. Algorithms that better utilizes all available data are expected to yield improved localization performance.\\
\indent In our previous work which was presented at the conference \cite{shen2016enhancement}, the problem of inferring the true vehicle positions, as well as the GNSS common biases from the pseudo-range measurements from a group of vehicles, is addressed by a Bayesian filtering approach. The aforementioned difficulties are solved by jointly estimating the common biases and vehicle positions using a Rao-Blackwellized particle filter (RBPF). In the RBPF, the correlation between common biases and vehicle positions is modeled implicitly through the diversity of the particles. As a result, there is no need for explicit data fusion. The effect of the multipath biases is mitigated through a detection-rejection method based on a statistical test. The particle filter structure allows multiple hypotheses with respect to the detection of multipath biases, thus making the detection more robust. In addition, the particle filter is flexible enough to handle road constraints in a straightforward manner by manipulating the particle weights according to the road constraints. It also fully exploits the temporal correlation through a prediction-update process, eliminating impossible configurations in the joint space of common biases and vehicle state variables, drastically reducing the estimation variance. The computational complexity varies linearly with the number of vehicles, which makes the proposed RBPF both effective and efficient.\\
\indent This paper expands the initial results from the conference paper. More specifically, the performance of the algorithm in multipath environments with signal blockage is shown through simulations based on 3-D Ray Tracing method for multipath error. The performance of the RBPF under open sky conditions is validated through experiments, which validates the pseudo-range error model. The robustness of the algorithm with respect to signal blockage is also studied. One potential drawbacks of this CMM method is that the localization accuracy and robustness highly depend on the configuration of the available road constraints. The impact of road constraints on CMM is discussed in our more recent work \cite{shen2017impact}.
\section{Theory and Method}
\begin{figure}[htbp]
  \centering
  \includegraphics[width=3.5in]{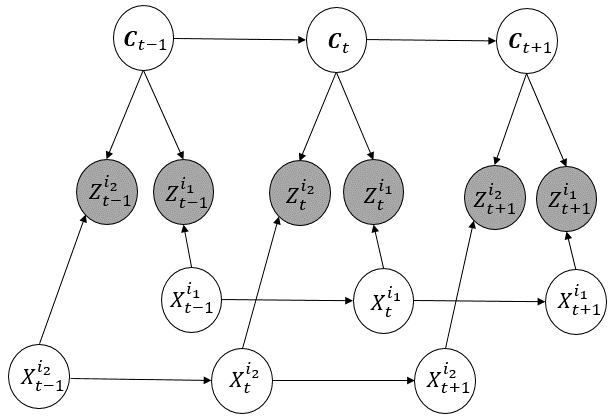}   
  \caption{DBN representation of the CMM problem in Fig.~\ref{localization_bias}}
  \label{bay_nets}
\end{figure}

In this section, the structure of the CMM problem is illustrated by a Dynamic Bayesian Network (DBN), which encodes the conditional independence that motivates the RBPF. The theoretical aspect of the RBPF is then introduced, and the prediction-update structure of the RBPF is shown in detail. \\
\indent Fig.~\ref{bay_nets} shows the DBN corresponding to the CMM problem that involves only two vehicles with index $i_1$ and $i_2$. $C$ represents the set of pseudo-range common biases, $X$ is the vehicle state vector including vehicle positions, and $Z$ represents the set of error corrupted pseudo-ranges, which is observed by the receivers. The subscript represents the time. The directed edges represent causal relationships between the node variables. For example, the pseudo-ranges $Z^{i_1}_t$ are determined by the vehicle state $X^{i_1}_t$ and the pseudo-range common biases $C_t$. As a result, $X^{i_1}_t$ is correlated with $C_t$ through the observation $Z^{i_1}_t$ and the the states of all the vehicles within the network are correlated with each other through their correlation with the pseudo-range common biases. This correlation is encoded by the paths between one vehicle state to another through the common biases nodes. If the common biases are conditioned on, then the paths would be blocked, indicating that the states of all the vehicles are independent with each other if the common biases are given. \\
\indent The following assumptions are made in this work:
\begin{enumerate}
 \item The non-common error of different vehicles are uncorrelated.
 \item The common biases vary slowly over time.
 \item The vertical positions of the vehicles can be obtained with reasonable accuracy from the digital map.
 \end{enumerate}
 
\indent Assuptions (1) - (3) are unrestrictive in the following senses. The first assumption is valid as long as the participating vehicles are not concentrated in the same area; otherwise, the multipath error may be correlated. It will be true for most rural areas and some urban areas. The second assumption is reasonable because the tropospheric and ionospheric delays, as the major components of the common biases, typically change very slowly over time \cite{kaplan2005understanding}. The third assumption is also reasonable, as the difference between the vertical positions recorded by the digital map and the ground truth can be considered equivalent measurement noise and accounted for by increasing the noise variance parameter.\\
\indent Assumption 1, together with the DBN representation Fig. \ref{bay_nets}, results in the following factorization of the joint posterior distribution of the pseudo-range common biases and the vehicle states conditioned on the pseudo-range observations:
\begin{equation}
\begin{aligned}
p(C^{1:N_s}_{1:t},X^{1:N_v}_{1:t}|Z_{1:t})=p(X^{1:N_v}_{1:t}|C^{1:N_s}_{1:t},Z_{1:t})p(C^{1:N_s}_{1:t}|Z_{1:t})\\=\prod^{N_v}_{i=1}p(X^{i}_{1:t}|C^{1:N_s}_{1:t},Z_{1:t})p(C^{1:N_s}_{1:t}|Z_{1:t}),
\end{aligned}
\end{equation}
where $N_s$ and $N_v$ are the number of satellites and vehicles respectively, and the superscript $1:N_s$ and $1:N_v$ are shorthands for the corresponding variables of all the satellites and all the vehicles. The subscript $1:t$ is the shorthand for the corresponding variables of all the time instances.\\
\indent The RBPF exploits this conditional independence property for efficient inference of the pseudo-range common biases and the vehicle states given the pseudo-range observations. The posterior distribution of the common biases $p(C^{1:N_s}_{1:t}|Z_{1:t})$ is estimated by particle filter, and the distributions of the vehicle states conditioned on the common biases $p(X^{i}_{1:t}|C^{1:N_s}_{1:t},Z_{1:t})$ are independent with each other and estimated by a set of EKFs whose dimension is the dimension of the state vector \cite{doucet2000rao}. The recursive prediction-update equations are presented as follow.\\ 
\subsection{Prediction of States}
With assumption 2, we model the time variation of the common biases as a first-order Gaussian-Markov process:
\begin{equation}
C^j_t=C^j_{t-1}+w^j_t{\Delta}t,   
\end{equation}
where $w^j_t{\sim}N(0,{\sigma}^2_c)$, with ${\sigma}^2_c$ denoting the variance of the common bias drift, ${\Delta}t$ is the length of the time interval between two successive updates of the states and $j=1,2,...,N_s$ is the index for satellites. \\
\indent Assumption 3 implies that only the horizontal positions and velocities need be modeled explicitly. Therefore, the state vector of the $i$th vehicle at time $t$ is\\
\begin{equation}
X^i_t=(\begin{array}{cccccc}
     x^i_t&{\dot x}^i_t&y^i_t&{\dot y}^i_t&b^i_t&{\dot b}^i_t\\
\end{array})^T,
\end{equation}
where $x^i_t$ and $y^i_t$ are the horizontal positions; ${\dot x}^i_t$ and ${\dot y}^i_t$ are the horizontal velocities; and $b^i_t$ and ${\dot b}^i_t$ are the receiver clock bias and drift, respectively.\\
\indent The mean is propagated by\\
\begin{equation}
\bar{X}^i_t=AX^i_{t-1},\textrm{with }
A=\left[
\begin{matrix}
B&0&0\\
0&B&0\\
0&0&B
\end{matrix}
\right],B=\left[
\begin{matrix}
1&{\Delta}t\\
0&1
\end{matrix}
\right],
\end{equation}

where $\bar{X}^i_t$ is the predicted mean.\\
\indent The associated covariance matrix is propagated by\\
\begin{equation}
\begin{aligned}
&\bar{\Sigma}^i_t=A{\Sigma}^i_{t-1}A^T+R_t,\textrm{with }
R_t=\left[
\begin{matrix}
R_x&0&0\\
0&R_y&0\\
0&0&R_b
\end{matrix}
\right],\\
&R_x=\left[
\begin{matrix}
\frac{\sigma^2_{ax}{\Delta}t^4}{4}&\frac{\sigma^2_{ax}{\Delta}t^3}{2}\\
\frac{\sigma^2_{ax}{\Delta}t^3}{2}&\sigma^2_{ax}{\Delta}t^2
\end{matrix}
\right],
R_y=\left[
\begin{matrix}
\frac{\sigma^2_{ay}{\Delta}t^4}{4}&\frac{\sigma^2_{ay}{\Delta}t^3}{2}\\
\frac{\sigma^2_{ay}{\Delta}t^3}{2}&\sigma^2_{ay}{\Delta}t^2
\end{matrix}
\right],\\
&R_b=\left[
\begin{matrix}
\frac{\sigma^2_{d}{\Delta}t^4}{4}+\sigma^2_{b}{\Delta}t^2&\frac{\sigma^2_{d}{\Delta}t^3}{2}\\
\frac{\sigma^2_{d}{\Delta}t^3}{2}&\sigma^2_{d}{\Delta}t^2
\end{matrix}
\right],
\end{aligned}
\end{equation}
where ${\Sigma}^i_{t-1}$ is the covariance matrix of the state vector; $\bar{\Sigma}^i_t$ is the predicted covariance matrix; $\sigma^2_{ax}$ and $\sigma^2_{ay}$ are the variances of the horizontal accelerations; and $\sigma^2_{b}$ and $\sigma^2_{d}$ are the variances of the clock bias and drift time derivatives \cite{giremus2007particle}, which are assumed to be uncorrelated in the derivation.
\subsection{Multipath Rejection and Measurement Update}
The pseudo-range measurement model between satellite $j$ and vehicle $i$ is\\
\begin{equation}
Z^{j,i}_t={\Vert}p^i_t-s^j_t{\Vert}+C^j_t+b^i_t+\lambda^{j,i}_tm^i_t+v^i_t,
\end{equation}
where $p^i_t$ is the position of the vehicle and $s^j_t$ is the satellite position. $m^i_t$ is the potential multipath bias and $v^i_t{\sim}N(0,{\sigma}^2_z)$ is the receiver noise, which is assumed to be white. $\lambda^{j,i}_t$ is a binary indicator variable that is to be determined through a $\chi^2$ test to indicate the presence of multipath bias.\\
\indent In the absence of multipath biases, the predicted mean of the pseudo-range measurement will be
\begin{equation}
\bar{Z}^{j,i}_t={\Vert}p^i_t-s^j_t{\Vert}+C^j_t+b^i_t
\end{equation}
\indent The difference between the actual pseudo-range measurement and the predicted mean will obey a Gaussian distribution, and the Mahalanobis distance of this random variable will obey the $\chi^2$ distribution with one degree of freedom:\\
\begin{equation}
D^2_{j,i}=(Z^{j,i}_t-\bar{Z}^{j,i}_t)^TP^{-1}_{j,i}(Z^{j,i}_t-\bar{Z}^{j,i}_t){\sim}\chi^2_1
\end{equation}
\begin{equation}
P_{j,i}=H_{j,i}\Sigma^i_{xy}H^T_{j,i}+\sigma^2_z, H_{j,i}=\frac{{\partial}Z^{j,i}_t}{\partial(x^i_t,y^i_t)},
\end{equation}
where $\Sigma^i_{xy}$ is the submatrix of the covariance matrix representing the uncertainty of the horizontal position, and $H_{j,i}$ is the Jacobian of the measurement function with respect to the horizontal position, which projects the uncertainty of the position space to the range space.\\
\indent The indicator variable is determined by\\
\begin{equation}
\lambda^{j,i}_t=\left\{
 \begin{array}{rcl}
 0       &      & D^2_{j,i}{\leq}F^{-1}(\alpha_1)\\
 1       &      & D^2_{j,i}{\geq}F^{-1}(\alpha_2){\Vert}u_{j,i}{\leq}\frac{F(D^2_{j,i})-\alpha_1}{\alpha_2-\alpha_1}
 \end{array} \right.,
\end{equation}
where $F$ is the Cumulative Distribution Function (CDF) of the $\chi^2_1$ distribution. $\alpha_1$ , $\alpha_2$ (with $\alpha_1<\alpha_2$) are the confidence levels for the rejection and acceptance of the multipath presence hypothesis, respectively. $u_{j,i}$ is a random number generated according to the uniform distribution on $[0,1]$, $\Vert$ is the logical ``or".\\
\indent The choices of $\alpha_1$ and $\alpha_2$ determine the aggressiveness to reject outliers. The particle filter keeps multiple hypotheses with respect to the assumptions on multipath biases. Particles that make wrong hypotheses will be eliminated by applying the map constraints.\\
\indent The weights of the particles are calculated according to the importance sampling principal. The detailed mathematical derivation can be found in \cite{montemerlo2002fastslam}. For the pseudo-range measurement from the $j$th satellite, the weights of the particles are updated as follows\\
\begin{equation}
w^{[k]}_j=\left\{
 \begin{array}{rcl}
 w^{[k]}_{j-1}\frac{1}{2\pi P_{j,i}}exp(-\frac{1}{2}D^2_{j,i})       &      & \lambda^{j,i}_t=0\\
 w^{[k]}_{j-1}\frac{1}{2\pi P_{j,i}}exp(-\frac{1}{2}F^{-1}(\alpha_3))       &      & \lambda^{j,i}_t=1
 \end{array} \right.,
\end{equation}
where $\alpha_3$ is a parameter that can be tuned depending on the environment and frequency of multipath occurrence, and superscript $[k]$ is the index of particles.\\
\indent The vehicle states are then updated using all the pseudo-range measurements regarded as free of multipath biases, that is, with $\lambda^{j_n,i}_t=0, n=1,2...N$ by
\begin{equation}
X^i_t=\bar{X}^i_t+K^i_t(Z^i_t-\bar{Z}^i_t),
\end{equation}
where $\bar{X}^i_t$ and $\bar{Z}^i_t$ are calculated by Eq.~(4) and Eq.~(7), respectively. $Z^i_t$ is the actual measurement. The Kalman gain matrix $K^i_t$ is calculated as:
\begin{equation}
K^i_t=\bar{\Sigma}^i_t(\tilde{H}^i_t)^T(\tilde{H}^i_t\bar{\Sigma}^i_t\tilde{H}^i_t)^T+Q_t)
\end{equation}
\begin{equation}
\Sigma^i_t=(I-K^i_t\tilde{H}^i_t)\bar{\Sigma}^i_t
\end{equation}
\begin{equation}
\tilde{H}^i_t=\left[
\begin{matrix}
\frac{{\partial}Z^{j_1,i}}{{\partial}x^i_t}&0&\frac{{\partial}Z^{j_1,i}}{{\partial}y^i_t}&0&0&0\\
...&...&...&...&...&...\\
\frac{{\partial}Z^{j_N,i}}{{\partial}x^i_t}&0&\frac{{\partial}Z^{j_N,i}}{{\partial}y^i_t}&0&0&0
\end{matrix}
\right]_{N{\times}6}
\end{equation}
\begin{equation}
Q_t=\left[
\begin{matrix}
\sigma^2_z&0&0\\
...&...&...\\
0&0&\sigma^2_z
\end{matrix}
\right]_{N{\times}N},
\end{equation}
where $\bar{\Sigma}^i_t$ is calculated by Eq.~(5); $\tilde{H}^i_t$ is the measurement Jacobian for batch update; $I$ is the identity matrix.
\subsection{Applying Map Constraint}
After the EKF update for the vehicle states estimation and updating the weight of the particles for the common biases estimation, the positioning of each vehicle represented by a set of EKFs with different weights will still be biased as no correction has been applied to compensate the pseudo-range common biases. In order to correct the biases, the map constraint is used to further modify the particle weights such that those particles with vehicle configurations incompatible with the map constraint are assigned a low weight and will be eliminated with high probability during the resampling. In this paper, the particle weights are modified by\\
\begin{equation}
\begin{aligned}
w^{[k]}_i=w^{[k]}_{i-1}{\int}\varepsilon(x^i_t,y^i_t)p(x^i_t,y^i_t)dx^i_tdy^i_t, i=1,2...N_v,\\
\textrm{with }
\varepsilon(x^i_t,y^i_t)=\left\{
 \begin{array}{rcl}
 1       &      & (x^i_t,y^i_t) \mbox{on lane}\\
 0       &      & (x^i_t,y^i_t) \mbox{out of lane}
 \end{array} \right.,
 \end{aligned}
\end{equation}
where $p(x^i_t,y^i_t)$ is the joint Gaussian distribution drawn from the EKF.\\
\indent The integral in Eq.~(17) is difficult to calculate analytically due to the potentially complicated geometry. Therefore, it is again calculated by Monte Carlo Integration, where the proposal distribution is $p(x^i_t,y^i_t)$ and the importance weight is $\varepsilon(x^i_t,y^i_t)$.
\begin{equation}
{\int}\varepsilon(x^i_t,y^i_t)p(x^i_t,y^i_t)dx^i_tdy^i_t\approx \frac{1}{N_m} \sum^{N_m}_{l=1}\varepsilon(x^{i,[l]}_t,y^{i,[l]}_t),
\end{equation}
where $(x^{i,[l]}_t,y^{i,[l]}_t), l=1,2...N_m$ is the set of samples drawn from the distribution $p(x^i_t,y^i_t)$, and $N_m$ is the size of the sample set.
\indent The pseudo code of the proposed RBPF is shown as follows:\\
  $(C^{[k]}_t,X^{[k]}_t,w^{[k]}_t)=RBPF(C^{[k]}_{t-1},C^{[k]}_{t-1},C^{[k]}_{t-1})$
\begin{enumerate}
 \item Predict $C_t$ and $X_t$ according to Eqs.~(2, 4)\\
 for vehicle $i=1:N_v$
 \item Determine the indicator variable according to Eq.~(10)
 \item Calculate particle weights and update $X_t$ according to Eqs.~(11, 12)
 \item Modify particle weights according to Eq.~(17)
 \item Resample\\
  end RBPF
 \end{enumerate}
\subsection{Computational Complexity}
The computational complexity of the RBPF is linear in the number of particles. For each particle at each time instance, a set of $N_v$ EKFs with dimensions equal to 6 have to be updated. As a result, the computational complexity is $O(N_pN_v)$, which is also linear in the number of vehicles. The linear complexity with respect to the number of connected vehicles is an attractive property as the potential number of vehicles can be hundreds. This prohibits the use of many filtering schemes such as Kalman filters, which has quadratic complexity, and particle filter without Rao-Blackwellization, which has exponential complexity.\\ 
\indent The number of particles $N_p$ has an effect on the estimation accuracy and robustness of the RBPF. The minimal $N_p$ to ensure robustness depends on the number of visible satellites and increases exponentially with the number of satellites. Nonetheless, this exponential growth does not pose a computational difficulty in practice because the number of visible satellites is always bounded. We show that the required number of particles for an accurate localization can be quite small ($O(10)$) in the next section.  

\section{Simulation Results and Discussions}
In this section, simulation results are presented and discussed. The simulation scenario is described first. The improvements in localization accuracy and the common biases estimation are illustrated by comparing the RBPF with algorithms proposed in Rohani et al. \cite{rohani2016novel}.
\subsection{Simulation Scenario and Error Models}
\begin{figure}[htbp]
  \centering
  \includegraphics[width=0.6\columnwidth]{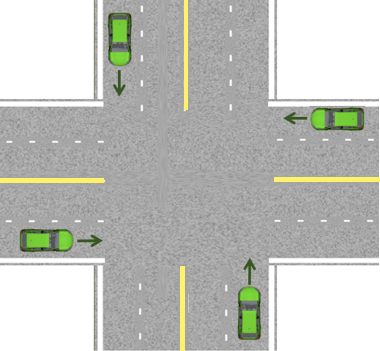}
  \caption{Intersection used for CMM}
  \label{f2}
\end{figure}

The configuration of the simulated scenario is shown in Fig.~\ref{f2}, where four vehicles are traveling in each lane of two orthogonal roads, respectively. The width of each lane is 3.5 \mbox{ }m and all the four vehicles are traveling on the center of the lanes. The boundary of the roads are represented by the white solid lines, outside which the positioning is considered as violating the road constraints. The performance of the proposed algorithm is illustrated through comparison with the CMM algorithm proposed in Rohani et al. \cite{rohani2016novel}. In their approach, the common correction of the vehicle positions is searched in the position space using a particle-based approach by applying map constraints. Due to the uncertainty caused by the non-common error, the map constraints are blurred to avoid over-convergence. This approach is not a Bayesian filtering approach as the estimation is not updated according to the Bayesian rule and the estimation at each time instance does not require any historical information.

\indent Three CMM algorithms are compared. The first algorithm is the aforementioned one proposed by Rohani et al. (referred to as the static method); the second algorithm is a smoothed version of the first one, where the GNSS positioning is smoothed by a Kalman filter before implementing CMM. The third algorithm is the proposed RBPF.\\
\indent The simulation parameters appear in Table 1， where $x$ and $y$ are local coordinates alone and transversal to the lane on which the vehicle travels. The simulation uses 200 particles. In the RBPF, the initial common biases are the true common biases corrupted by white noise, with variance $\sigma^2_n=0.25\mbox{ }m^2$. In the simulation, the clock error is not included.

\begin{table}[ht] 
\caption{Simulation parameters} 
\centering      
\begin{tabular}{c c c | c c c}  

\hline\hline                        
Parameter & Value & Unit & Parameter & Value & Unit \\ [0.5ex] 
\hline                    

$N_m$ & 100 & /&$\sigma_{ax}$ & 1 & $m/s^2$ \\
$N_s$ & 6 & / &$\sigma_{ay}$ & 0.1 & $m/s^2$ \\
$\alpha_1$ & 0.95 & / & $\sigma_b$ & 1 & $m/s$ \\
$\alpha_2$ & 1 & / & $\sigma_c$ & 0.1 & $m/s$ \\
$\alpha_3$ & 0.99 & / & $\sigma_d$ & 1 & $m/s^2$ \\
$\Delta t$ & 0.1 & $s$ & $\sigma_z$ & 1 & $m$ \\ 

\hline     
\end{tabular}
\label{table:nonlin}  
\end{table}

\indent The performances of these three algorithms under two measurement noise models are simulated. The first noise model simulates common biases and uncorrelated white noise with variance $\sigma^2_z$; the second noise model simulates common biases, uncorrelated white noise with variance $\sigma^2_z$ and multipath biases. \\
\indent The common biases and the satellite constellation are emulated using the GPSoft Satellite Navigation Toolbox \cite{SatNav}. This toolbox emulates the constellation using Keplerian orbital parameters and the satellite orbit error is neglected. The common biases are generated according to its empirical ionospheric and tropospheric error models. The multipath signals are simulated by Ray-tracing method given a 3-D digital map of the environment. The delay-lock loop (DLL) with half chip length correlation function is simulated to mimic the code tracking mechanism of the GNSS receiver. This DLL causes the code multipath error to oscillate as a function of the receiver location. More details about the code tracking mechanism can be found in \cite{braasch1999gps}. The reflected signal strength and phase difference relative to those of the direct path are required to simulate the DLL mechanism if both of the two signals exist. Therefore, following simplifications are made with respect to the multipath simulation: 
\begin{enumerate}
 \item The amplitude of the signal is reduced by a half upon reflection, that is, the reflection coefficients of all the building surfaces are assumed to be a constant value $0.5$. 
 \item Signals that have been reflected twice are not considered for multipath contribution due to their negligible signal strength.
 \item If the direct path does not exist, the received signal, no matter exists or not, is not used to calculate the corresponding pseudo-range because of the low signal to noise ratio. 
 \end{enumerate}
\indent Based on the first simplification, the GPS signal including the multipath signal can be written as
\begin{equation}
s(t)=a_0e^{-I(\omega_0t+\phi_0)}[x(t-t_0)+\frac{1}{2}e^{-I\phi_d}x(t-t_0-t_d)],
\end{equation}
where $a_0$ is the amplitude of the direct path, $I=\sqrt{-1}$ is the imaginary unit, $\omega_0$ is the angular frequency of the GPS carrier wave, $\phi_0$ is the received carrier phase of the direct path, $x(t)$ is the complex wave form of the transmitted signal, $t_0$ is the time for the signal to propagate from the satellite to the receiver through the direct path, $\phi_d$  and $t_d$ are the multipath phase delay and propagation time delay relative to the direct path, respectively.\\
\indent The composite wave form inside the square bracket in Eq.~(19) is correlated with the receiver generated wave forms $x(t-t_p+t_c)$ and $x(t-t_p-t_c)$, where $t_p$ is the multipath-corrupted signal propagation time to be determined by the DLL, and $t_c$ is the time for light to travel a half chip length. The results of these correlations are two functions of $t_p$, which are defined as early and late correlator function, denoted as $R_E(t_p)$ and $R_L(t_p)$, respectively. Their difference $D(t_p)=R_E(t_p)-R_L(t_p)$ is defined as discriminator function. In the absence of the multipath signal, the propagation time of the direct path will be the unique zero of the discriminator function within a neighbor of radius $2t_c$, that is, $D(t_0)=0$. Hence the receiver DLL determines the signal propagation time by finding the zero of the discriminator function, that is, $t_p=\mathop{arg}\limits_{t}\left\{D(t)=0\right\}$.\\
\indent For the single multipath-corrupted signal represented by Eq.~(19), simple analytic formula for the pseudo-range multipath error can be obtained as
\begin{equation}
\rho=c(t_p-t_0)=\frac{cos\phi_d}{2+cos\phi_d}t_d,
\end{equation}
where $c$ is the nominal speed of light in a vacuum. The multipath time delay $t_d$ is determined by
\begin{equation}
t_d=\frac{d_{d}-d_{m}}{c},
\end{equation}
where $d_{d}$ and $d_{m}$ is the path length through the direct path and the multipath, respectively, which can be obtained from the Ray-tracing method.\\ 
\indent The multipath phase delay $\phi_d$ is determined by
\begin{equation}
\phi_d=\omega_0t_d+\pi,
\end{equation}
where the $\pi$ is to account for half-wave loss upon reflection. 
\indent The aim of this multipath model is not to provide a faithful deterministic multipath simulation but rather to capture the major features of this multipath error signal that determine how well the localization algorithms might perform. Those features should include the approximate order of magnitude, the location dependence and the time variation characteristics of the multipath signal, which are expected to be reflected by this simplified model.\\
\indent As the wavelength of the GPS signal carrier wave ($\lambda=0.19\mbox{ }m$ for GPS L1 signal) is typically much smaller than the distances between the receiver to the buildings, the multipath phase delay is expected to be sensitive to the receiver motion. As a result, the multipath error should oscillate rapidly as the phase delay changes. This oscillating error may have similar effect as an additive white noise, which increases the apparent noise covariance of the observed pseudo-ranges. In order to account for this effect, the noise covariance used by the Kalman filter is estimated from the innovation. One of the unbiased estimations using $k$ steps previous innovations is
\begin{equation}
\tilde{Q_t}=\frac{1}{k-1}\sum_{j=t-k+1}^{t}(Z_j-\bar{Z_j})(Z_j-\bar{Z_j})^T-H_t\bar{\Sigma}_tH_t^T,
\end{equation}
where $Z$, $\bar{\Sigma}$ and $H$ have been defined in Eq.~(12) and Eq.~(13), with the vehicle identification omitted.\\
\indent Then the diagonal noise covariance matrix is formed such that
\begin{equation}
Q_t(n,n)=max(\tilde{Q_t}(n,n),\sigma^2_z), n=1,2,...N,
\end{equation}
where $(n,n)$ denotes the n-th diagonal element of a square matrix.

\begin{figure}[htbp]
  \centering
  \includegraphics[width=2.5in]{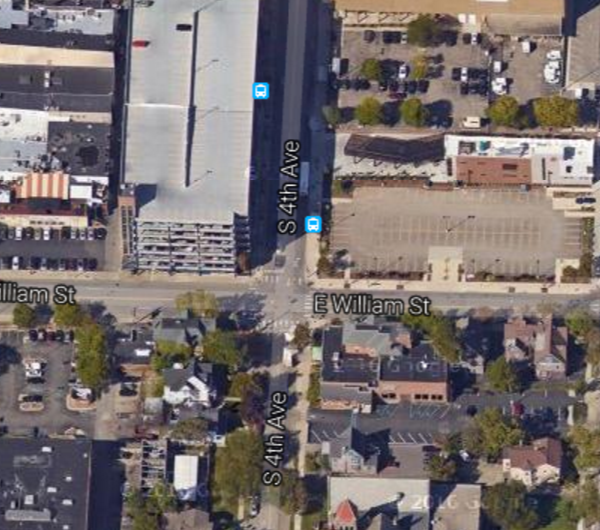}
  \caption{Intersection between South 4th Avenue and East William Street in Ann Arbor used for multipath simulation (Google Earth)}
  \label{f3}
\end{figure}

\indent The 3-D map around the intersection between South 4th Avenue and East William Street in Ann Arbor is obtained from Google Earth, shown in Fig.~\ref{f3}. The surrounding buildings are modeled as rectangular blocks. The geometric quantities of the map, including the lane widths and building locations and dimensions, are measured using the ``Ruler" tool provided in Google Earth Pro.

\subsection{Localization Results}
\begin{figure}[htbp]
  \centering
  \includegraphics[width=3.5in]{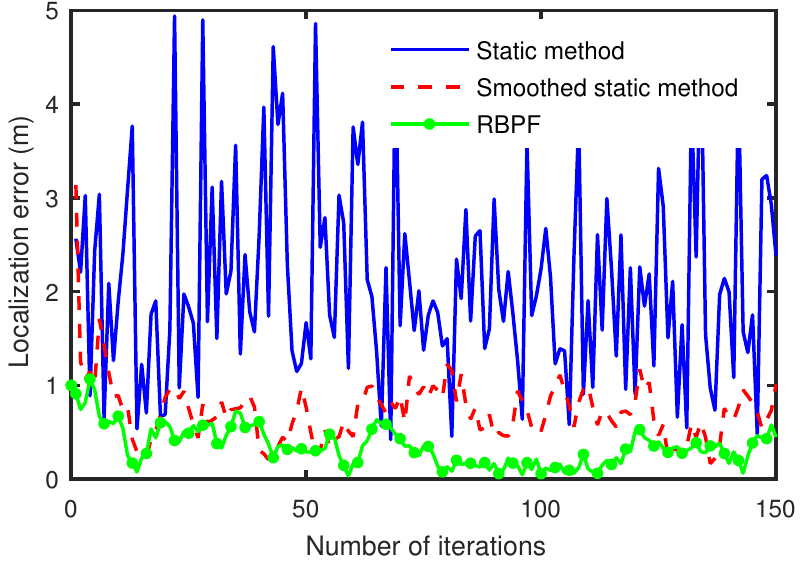}
  \caption{Horizontal position errors with common biases and white noise}
  \label{f4}
\end{figure}

\begin{figure}[htbp]
  \centering
  \includegraphics[width=3.5in]{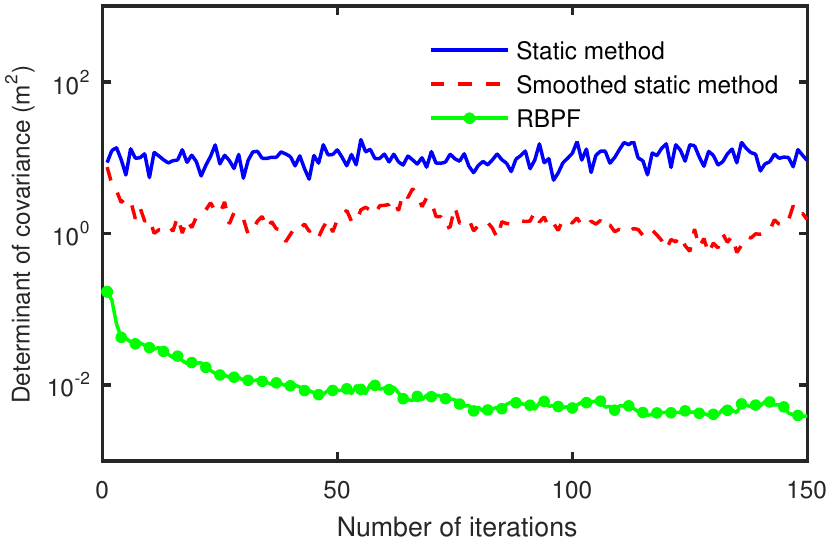}
  \caption{Determinant of horizontal position covariance with  common biases and white noise}
  \label{f5}
\end{figure}

\begin{figure}[htbp]
  \centering
  \includegraphics[width=3.5in]{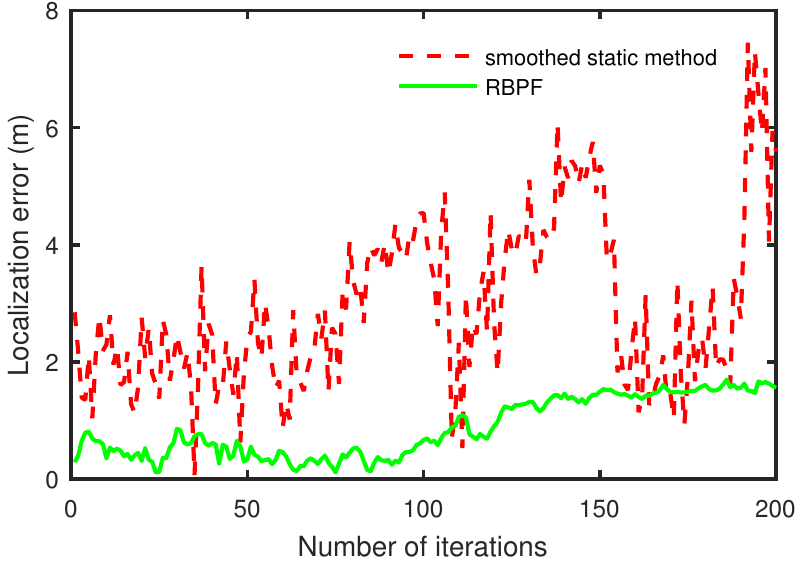}
  \caption{Horizontal position error with common biases and white noise+multipath error}
  \label{f6}
\end{figure}

The horizontal position error and the associated covariance of one of the four vehicles using the three described algorithms are shown in Fig.~\ref{f4}~-~\ref{f6}.\\
\indent Fig.~\ref{f4} shows that the localization error using the static method is much larger and noisier than either that using the smoothed static method or the proposed RBPF. This is to be expected, as the white noise results in non-common error, which is not filtered in the static method. As the smoothed algorithm filters out this non-common error, it  outperforms the static method. Nevertheless, the RBPF outperforms both of them.

\begin{figure*}[htbp] 
\centering
\includegraphics[width=3.3in]{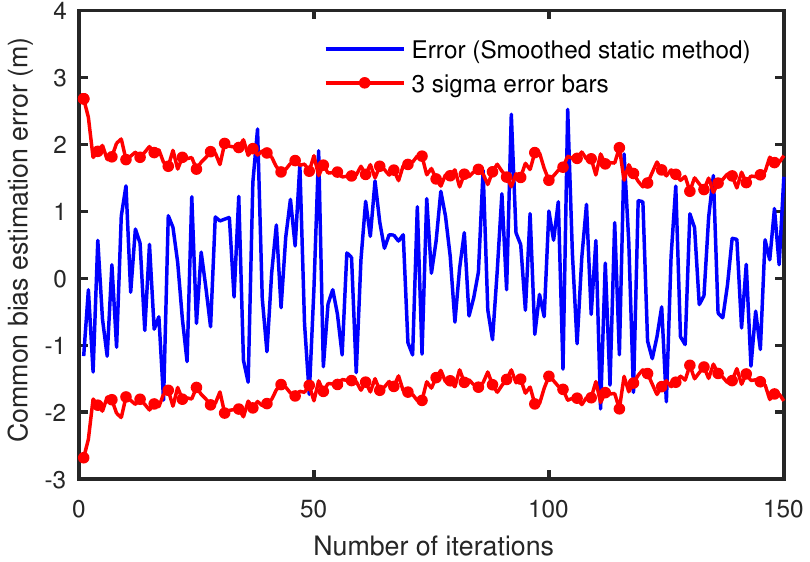}
\includegraphics[width=3.3in]{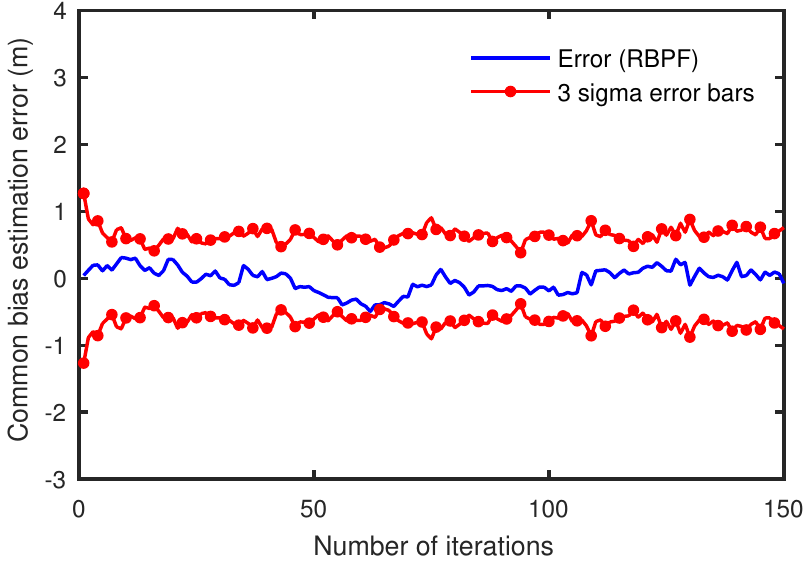}
\caption{Common bias estimation error}
%
%
%
\label{f7}
\end{figure*}

\indent Another benefit of the proposed RBPF is that it significantly outperforms the other two algorithms in terms of estimation covariance (see Fig.~\ref{f5}), because all the particles have a smooth estimation of the common error. In contrast, the other two methods search for all the compatible corrections for the common error at each instant. This search is both unnecessary and ineffective because most of the corrections, though compatible with the current map constraints, would be eliminated if the previous measurements were also considered. In other words, the temporal correlation of the common biases is not exploited. In contrast, the proposed RBPF keeps track of the most probable common biases. The common biases of small probability are eliminated through resampling, and the time correlation of the common biases is enforced by Eq.~(2). Thus, the estimation covariance turns out to be much smaller than that of the other two algorithms.\\
\indent Fig.~\ref{f6} shows the localization error using the smoothed static method and the RBPF in the presence of multipath error and signal blockage. The static method is not able to give localization results at all the time instants due to the signal blockage, so it is not compared. As the number of iteration increases, the vehicles approaches the intersection, where the multipath reflection and signal blockage from the high structure are severe. As a result, the localization error increases. The performance of the smoothed static method degrades severely while that of the RBPF does not degrade as much. This result is also expected as the smoothed static method uses the positions instead of the raw pseudo-range measurements to do the CMM. In the presence of signal blockage, different vehicles might use different sets of satellites for their ego-localization, which results in an unknown non-common position bias in addition to the multipath error. This additional non-common bias makes CMM more difficult. In contrast, the RBPF is formulated based on the raw pseudo-range measurements, which is more flexible to handle the signal blockage issue.\\
\indent Table 2 shows the mean of the localization error of these three methods, where $\pm$ is followed by the $3-\sigma$ confidence interval.\\
\begin{table}[ht] 
\caption{Mean localization error (m)} 
\centering      
\begin{tabular}{c c c c}  
\hline\hline                        
$\mbox{ }$ &Static &Smoothed static&RBPF\\ [0.3ex] 
\hline                    
Multipath free & 2.37 & 0.79$\pm$0.15 &0.45$\pm$0.18  \\   
Multipath included & - & 2.92&0.87\\ [0.3ex]
\hline     
\end{tabular}

\end{table}
\subsection{Estimation Results of Common Biases}

The left and right figures in Fig.~\ref{f7} show the common bias estimation error corresponding to one of the satellites in the multipath-free case using the smoothed static method and the proposed RBPF, respectively.\\
\indent Both of the two estimation errors are consistent with their covariance, while the estimation from the RBPF is more accurate and effective than that using the smoothed static method. In addition, since the RBPF estimates the common biases through filtering while the smoothed static method uses only instantaneous measurements, the estimation using the latter is noisier than that using the former.
\subsection{Computational Complexity}

\begin{figure}[htbp]
  \centering
  \includegraphics[width=3.2in]{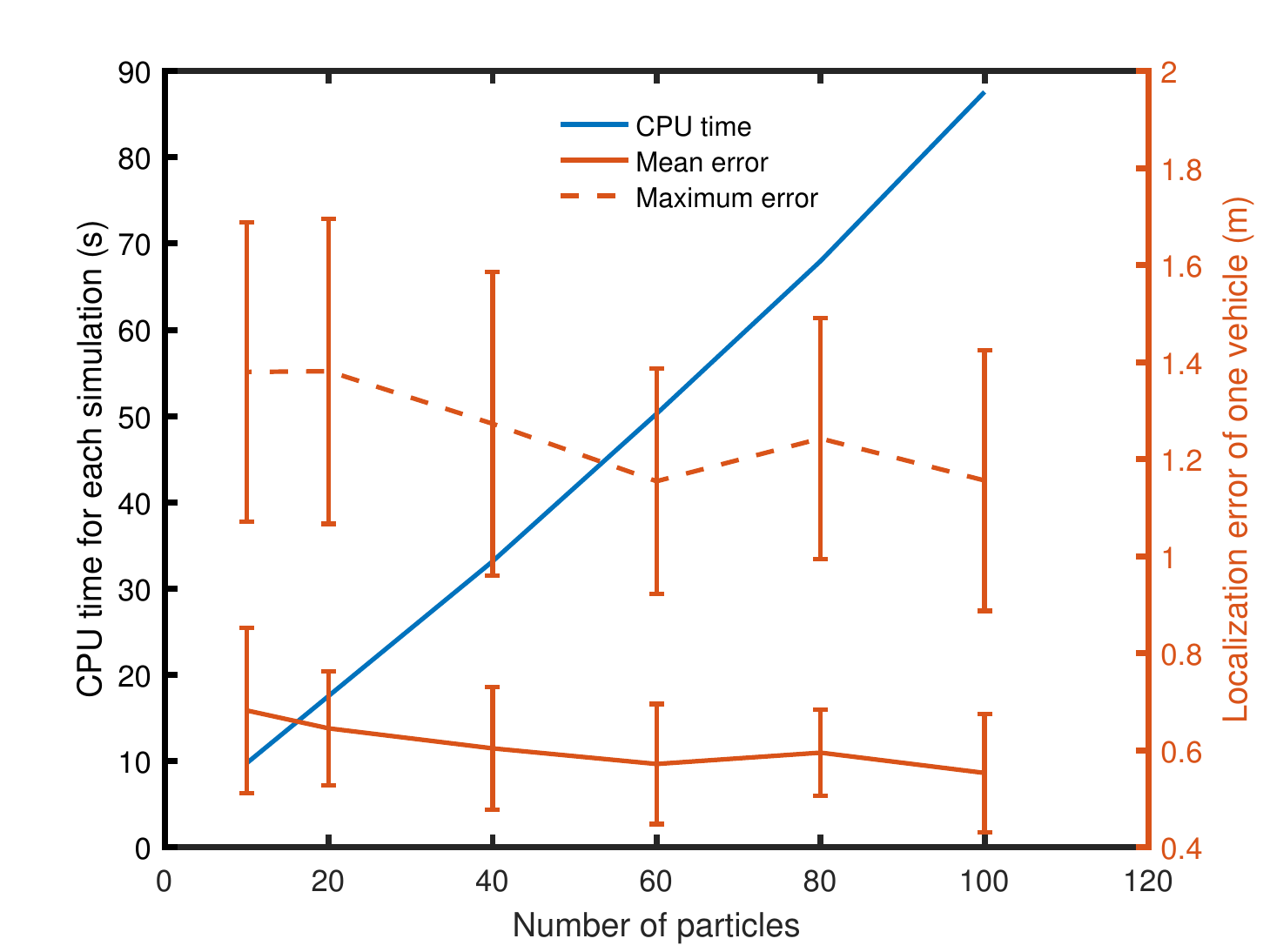}
  \caption{Computational complexity, mean and maximum localization error with respect to the number of particles used in the RBPF}
  \label{com_error}
\end{figure}

Fig.~\ref{com_error} shows the effect of the number of particles on the computation time and localization error. We ran 10 simulations for each point with 4 vehicles and 6 satellites under multipath free conditions on MATLAB 2016a with an Intel i-7 6500U processor. There are 300 time instances in each simulation and the time step is 0.1 s, which is equivalent to 30 s simulation time. As the number of particles increases, the computation time grows linearly and the localization error decreases. Even if the number of particles is small, the localization accuracy and robustness do not degrade too much.

\section{Experiment validation}
In this section, experimental results conducted at North Campus parking lot 90 in the University of Michigan are presented. First, the experiment scenario is described. Then the raw measurement data collected from the low-cost GPS receivers is processed and used to validate the presumed GPS pseudo-range error model presented in Section 3. After that, the CMM algorithms are applied on the raw data, and the CMM localization results are compared with each other and with ego localization results.
\subsection{Experiment Scenario Description}
\begin{figure}[htbp]
  \centering
  \includegraphics[width=2.5in]{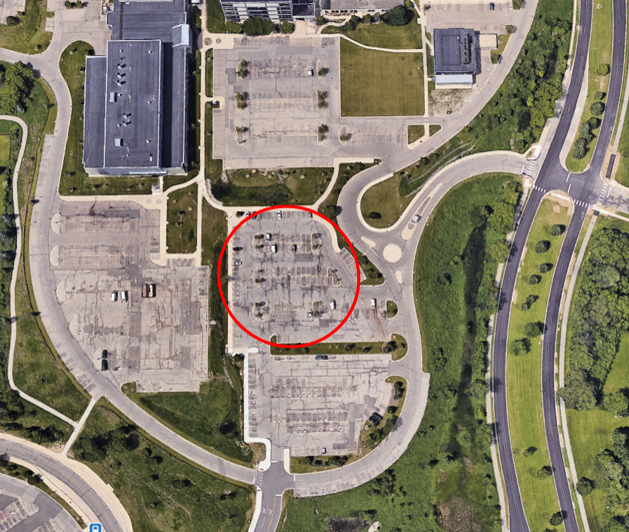}
  \caption{The surroundings of the experiment place and the parking lot (encircled by the red circle) where the experiment was conducted, image from Google Earth}
  \label{f_exp}
\end{figure}
The experiment was conducted at open sky at the parking lot shown in Fig.~\ref{f_exp}. We drew straight lines on the ground as virtual road boundaries according to the sketch Fig.~\ref{f2}. Four u-blox EVK-6T receivers were placed 2 m (about half a lane width) away from the corresponding road boundaries, with their configuration the same as that of the four vehicles shown in Fig.~\ref{f2}. The distance of the receivers from the road boundaries could affect the compactness of the CMM positioning. As the distance decreases, the road constraints become tight and so does the positioning covariance. In the limiting case that the distance goes to zero, the positioning covariance reaches its non-zero lower bound. In the limiting case that the distance goes to infinity, the CMM becomes almost equivalent to ego-localization.\\
\indent A Novatel DL-4 RTK GPS, with horizontal localization error less than 2 cm, was used to measure the position of the u-blox receiver on the upper left side, hereafter referred as the host receiver. The antenna of the host receiver was mounted on the steel top of a vehicle and those of the other three receivers were placed on the ground. This placement eliminated the multipath interference from the ground reflection. The raw measurements, including the pseudo-ranges, the corresponding satellite identifications and the time stamps,  were logged at a sampling rate of 5 Hz, while the four receivers were kept static. The satellite ephemeris broadcast by the satellites were logged to calculate the satellite positions.
\begin{figure}[htbp]
  \centering
  \includegraphics[width=3.5in]{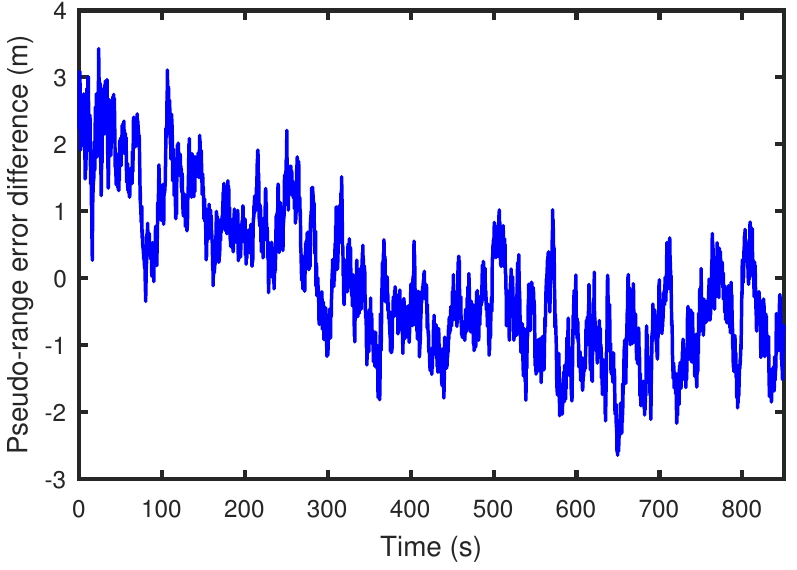}
  \caption{Pseudo-range error, including the slowly varying atmospheric delay (several minutes), the multipath error (about one minutes) and the white receiver noise (less than seconds)}
  \label{f8}
\end{figure}
\subsection{Validation of pseudo-range error model}
\begin{figure}[htbp]
  \centering
  \includegraphics[width=3.5in]{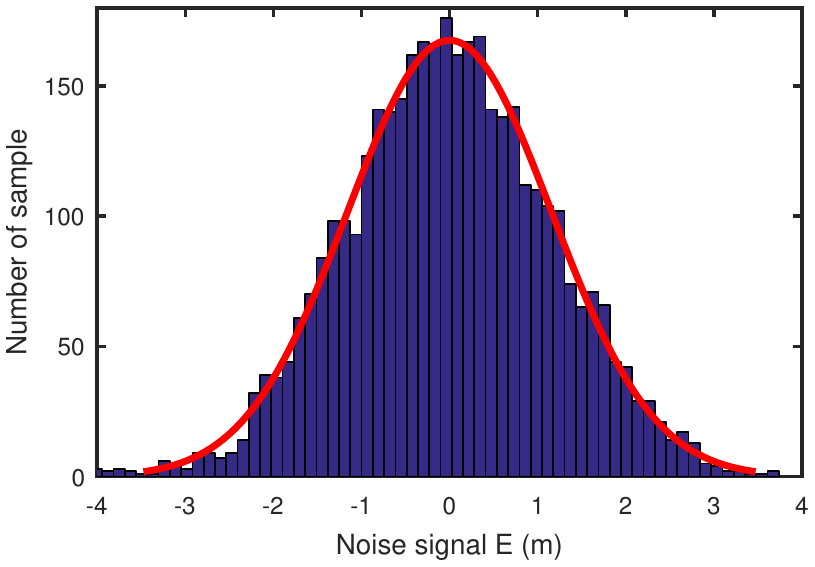}
  \caption{Histogram of the composite receiver noise signal}
  \label{f10}
\end{figure}

In Section 3, it is assumed that the range error is comprised of slowly varying common biases, non-common white Gaussian noise and clock biases at open sky. In this section, these assumptions are examined by processing the raw pseudo-range measurements.\\
\indent Fig.~\ref{f8} shows the composite pseudo-range signal by differencing the pseudo-range corresponding to two satellites collected by two receivers. This signal includes the variation of the atmospheric delay, the multipath error and the receiver noise. \\
\indent The histogram of the receiver noise signal is plotted in Fig.~\ref{f10}. It shows that the histogram can be reasonably fitted by a Gaussian distribution.

\begin{figure}[htbp]
  \centering
  \includegraphics[width=2.5in]{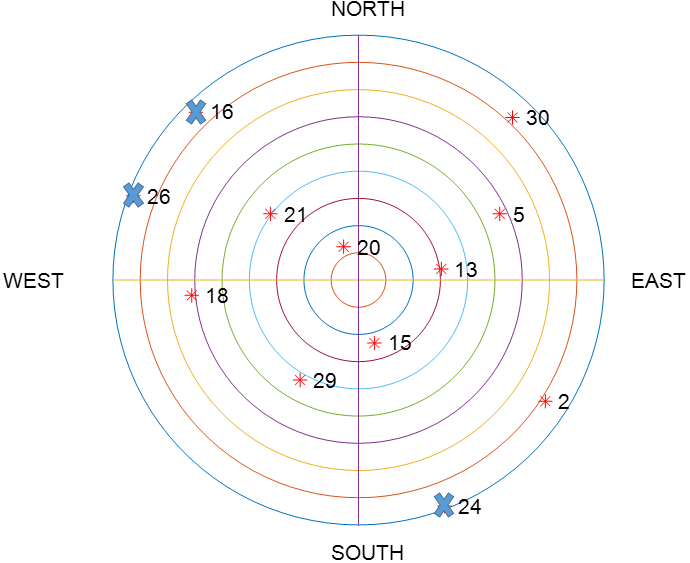}
  \caption{Skyplot of the GPS constellation at University of Michigan North Campus during the experiment. Satellite identification number as well as their availability are marked.}
  \label{f12}
\end{figure}

\subsection{Localization Results}
In order to validate the CMM algorithms in real world, a segment of the experiment data, collected from the Coordinated Universal Time (UTC) 20:56:40 to 21:01:40 on September 24th, 2016, is used for CMM. During this time period, the skyplot of the visible satellites at the experiment location according to the almanac is shown in Fig.~\ref{f12}.

\begin{figure*} \centering 
\subfigure[Open sky] { \label{fig:a} 
\includegraphics[width=0.75\columnwidth]{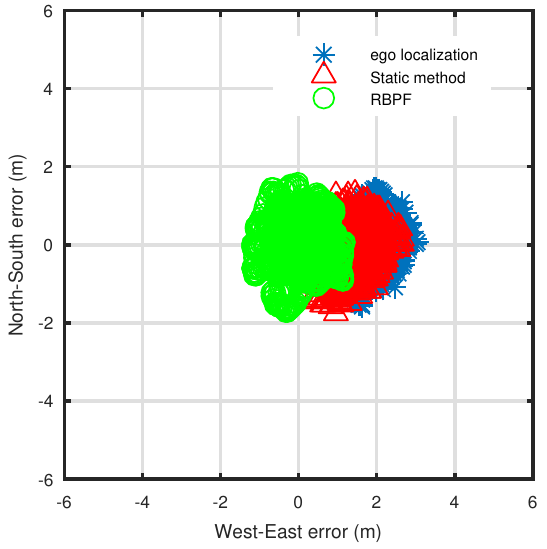} 
} 
\subfigure[With signal blockage] { \label{fig:b} 
\includegraphics[width=0.75\columnwidth]{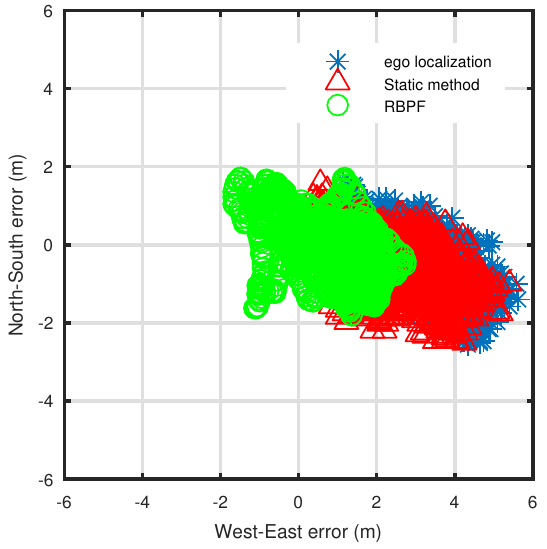} 
} 
\caption{Localization error using ego-localization, the static method and the RBPF on experimental data} 
\label{f13} 
\end{figure*}

\indent In Fig.~\ref{f12}, the three satellites denoted by the crosses, G16, G24 and G26, were not received due to blockage and low elevation angle. Those seven satellites of high elevation angle, that is, G5, G13, G15, G18, G20, G21 and G29, are used for localization. The corresponding horizontal dilution of precision (HDOP) is $2.79$. The HDOP indicates how much the error of the pseudo-range will result in the horizontal positioning error.\\
\indent The multipath error appears as a bias within a short time interval, which causes the localization results to be biased. This localization bias may result in over-confidence. The adaptive covariance approach used for localization in multipath environment is applied here. \\
\indent The simulation parameters used are the same as those listed in Table 1, except that the time interval between two successive time instants is 0.2 s; the number of satellites is 7; and the covariance of the receiver noise is estimated from the data. The initial values of the receiver position, clock bias and clock drift are obtained from the ego-localization results. The initial values of the receiver horizontal velocity are given as zeros. The initial values of the common bias are estimated from the difference between the RTK solution and the measured pseudo-ranges. \\
\indent The resulted localization error using ego-localization, the static method and the RBPF using 100 particles are shown in Fig.~\ref{f13}.a. It shows that the RBPF effectively eliminates the bias, while the static method partially eliminates the bias. Nevertheless, neither of these two algorithms is able to make effective reduction on the non-common error which is driven by the low frequency part of the receiver noise, as expected. \\
\indent Performance degradation in the presence of signal blockage is also of practical interest. In order to study the effect of signal blockage, the pseudo-range measurements from satellites G18 and G21 received by the host vehicle receiver are not used. The resulted localization error is shown in Fig.~\ref{f13}.b. In this case, the localization results given by the static method track the biased ego-localization results more closely. This degradation should be attribute to the use of different sets of satellites, which results in additional non-common error that is not modeled in the static method. In contrast, the localization results from the RBPF is almost unbiased, although the error is also amplified due to the absence of the two satellites, which increases the HDOP to 10.71. 
%

\section{Conclusions}
In this paper, a Rao-Blackwellized particle filter has been proposed for the simultaneous estimation of GNSS common biases and vehicles cooperative localization using map matching. The following conclusions can be drawn based on the simulation and experimental results:
\begin{enumerate}
 \item The proposed method fully exploits the temporal correlation of the common biases and vehicle positions through the prediction-update process such that the estimation covariance is reduced by at least two orders compared with previously proposed algorithms. 
 \item The proposed method almost entirely eliminates the slowing varying common localization bias, thus achieving a higher accuracy than both ego-localization and the previous CMM algorithm. Nevertheless, none of these three methods can effectively eliminate the low frequency part of the receiver noise error. The localization error with experiment at open sky is within 2 meters.
 \item The proposed method is more robust with respect to signal blockage than the previous CMM algorithm. With moderate signal blockage, the proposed method is still able to eliminate the common bias effectively, while the localization will be less accurate as the HDOP increases due to the loss of satellites. 
\end{enumerate}





%





\ifCLASSOPTIONcaptionsoff
  \newpage
\fi

\bibliography{ITS_bib}
\bibliographystyle{IEEEtran}

%

\begin{IEEEbiography}[{\includegraphics[width=1in,height=1.25in,clip,keepaspectratio]{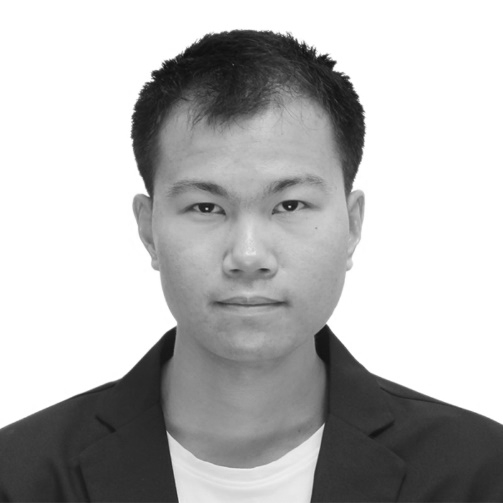}}]{Macheng Shen}
Macheng Shen received his B. S. degree in 2015 from Shanghai Jiao Tong University and his M. S. E. degree in 2016 from University of Michigan, Ann Arbor. He is currently a Ph. D. student in University of Michigan, Ann Arbor. His research interest includes connected vehicle localization and Bayesian filtering.   
\end{IEEEbiography}

\begin{IEEEbiography}[{\includegraphics[width=1in,height=1.25in,clip,keepaspectratio]{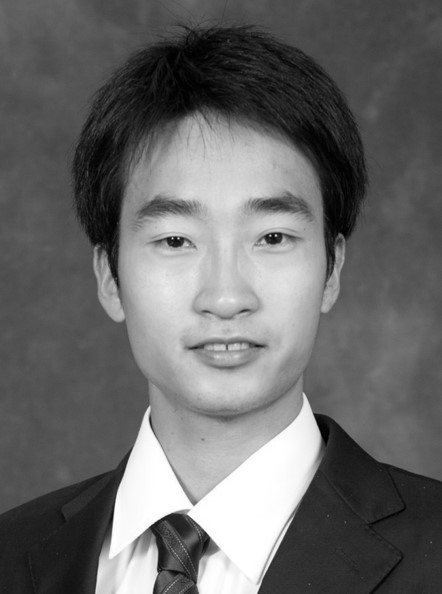}}]{Ding Zhao}
Ding Zhao received the Ph.D. degree in 2016 from the University of Michigan, Ann Arbor. He is currently a Research Fellow in the University of Michigan Transportation Research Institute. His research interest includes evaluation of connected and automated vehicles, vehicle dynamic control, driver behaviors modeling, and big data analysis
\end{IEEEbiography}


\begin{IEEEbiography}[{\includegraphics[width=1in,height=1.25in,clip,keepaspectratio]{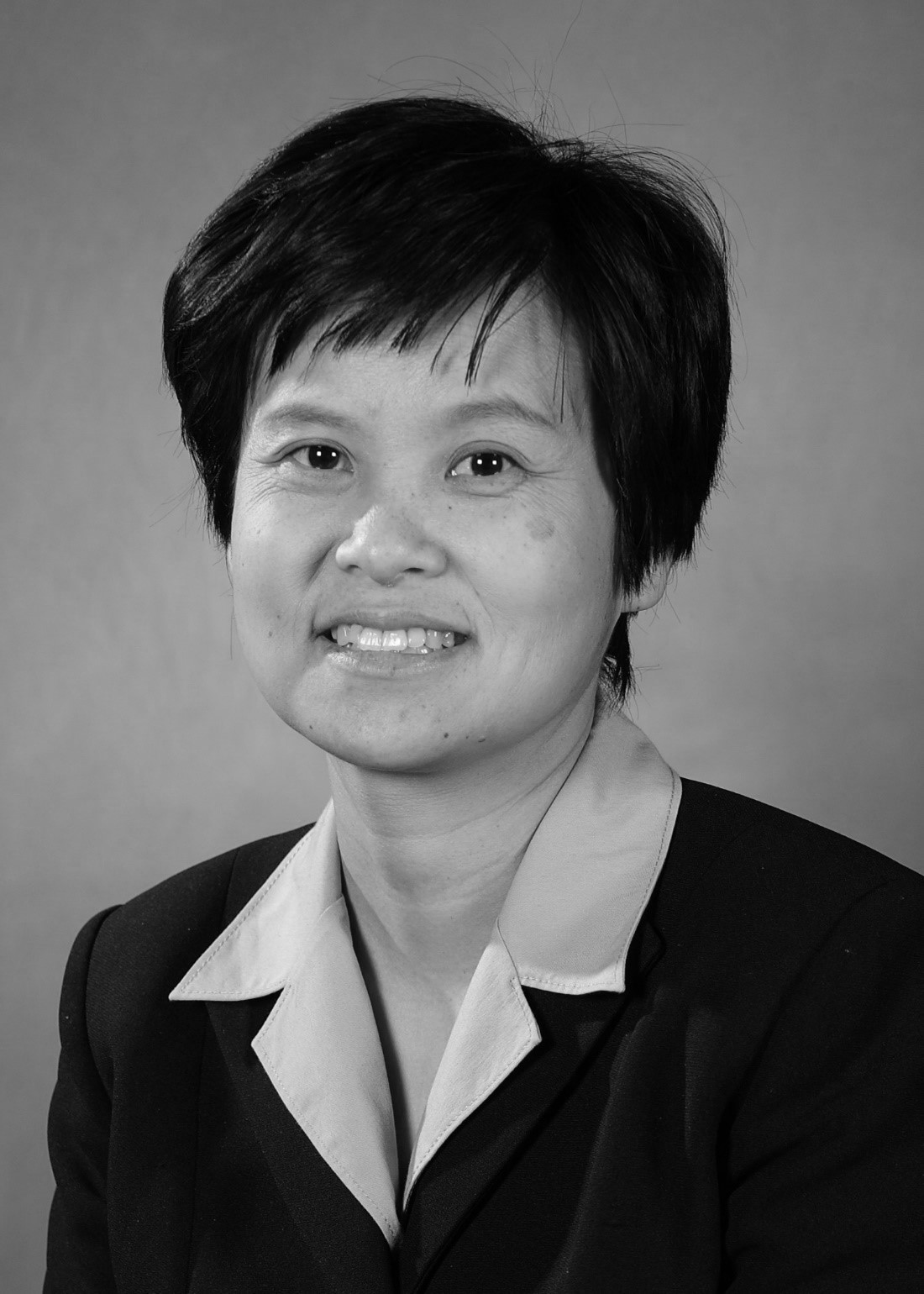}}]{Jing Sun}
Jing Sun received her Ph. D. degree from University of Southern California in 1989, and her B. S. and M. S. degrees from University of Science and Technology of China in 1982 and 1984 respectively. From 1989-1993, she was an assistant professor in Electrical and Computer Engineering Department, Wayne State University. She joined Ford Research Laboratory in 1993 where she worked in the Powertrain Control Systems Department. After spending almost 10 years in industry, she came back to academia and joined the faculty of the College of Engineering at the University of Michigan in 2003, where she is now Micheal G. Parsons Professor in the Department of Naval Architecture and Marine Engineering, with courtesy appointments in the Department of Electrical Engineering and Computer Science and Department of Mechanical Engineering. Her research interests include system and control theory and its applications to marine and automotive propulsion systems. She holds 39 US patents and has co-authored a textbook on Robust Adaptive Control. She is an IEEE Fellow and a recipient of the 2003 IEEE Control System Technology Award.
\end{IEEEbiography}

\begin{IEEEbiography}[{\includegraphics[width=1in,height=1.25in,clip,keepaspectratio]{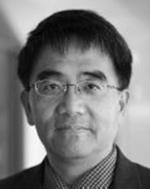}}]{Huei Peng}
Huei Peng received the Ph.D. degree from the University of California, Berkeley, CA, USA, in 1992. He is currently a Professor with the Department of Mechanical Engineering, University of Michigan, Ann Arbor, MI, USA. He is currently the U.S. Director of the Clean Energy Research Center—Clean Vehicle Consortium, which supports 29 research projects related to the development and analysis of clean vehicles in the U.S. and in China. He also leads an education project funded by the Department of Energy to develop ten undergraduate and graduate courses, including three laboratory courses focusing on transportation electrification. He has more than 200 technical publications, including 85 in refereed journals and transactions. His research interests include adaptive control and optimal control, with emphasis on their applications to vehicular and transportation systems. His current research focuses include design and control of hybrid vehicles and vehicle active safety systems.
\end{IEEEbiography}

\end{document}